# Analytical Boundary-Based Method for Diffraction Calculations

Eitam Luz[1,2], Er'el Granot[2] and Boris Malomed[1]

[1] Department of Physical Electronics, School of Electrical Engineering, Faculty of Engineering, Tel Aviv University, Tel Aviv 69978, Israel

[2] Department of Electrical and Electronic Engineering, Ariel University, Ariel, Israel

E-mail: eitamluz@tau.ac.il

## Abstract

We present a simple method for calculation of diffraction effects in a beam passing an aperture. It follows the well-known approach of Miyamoto and Wolf, but is simpler and does not lead to singularities. It is thus shown that in the near-field region, i.e., at short propagation distances, most results depend on values of the beam's field at the aperture's boundaries, making it possible to derive diffraction effects in the form of a simple contour integral over the boundaries. For a uniform, i.e., plane-wave incident beam, the contour integral predicts the diffraction effects exactly. Comparisons of the analytical method and full numerical solutions demonstrate highly accurate agreement between them.

## 1. Introduction

Diffraction is a ubiquitous phenomenon in wave dynamics (optics, acoustic, propagation of radio waves, etc.) [1, 2, 3]. In particular, diffraction occurs when a beam passes an aperture in a diffraction screen. The first (scalar) method for modeling diffraction was initiated by Huygens and developed by Fresnel [1] (since then known as the Fresnel paraxial diffraction integral), is that every point on the wavefront is regarded as a source of spherical waves, and the sum of these wavelets form the wavefront at any later time and distance [1, 2]. The second method, known as the boundary diffraction wave theory (BDWT), was initiated by Young [4]. It is based on the assumption that diffraction is an intermediate phenomenon between the unobstructed propagating beam (also known as the geometrical wave) and a scattered edge wave (known as the boundary diffraction wave) [2]. A new version of the BDWT technique for arbitrary beams was elaborated by Miyamoto and Wolf. They decomposed the surface integral in the Fresnel-Kirchhoff diffraction formula into a contour integration (boundary diffraction wave) and a geometrical wave [5].

The former method is much more commonly used in the diffraction analysis, while broad adoption of BDWT was impeded by its relative complexity. In particular, the contour integration in Ref. [5] makes it necessary to handle singularities at edges of the geometric shadow (i.e., the shadow in the absence of diffraction). Consequently, the method elaborated in Ref. [5] was implemented, thus far, only for Gaussian beams (except for the plane-wave and point-source scenarios) [6]. However, the BDWT method clearly offers a potential to make the numerical analysis more efficient, replacing 2D integrals by 1D counterparts, which may be computed much faster. Moreover, this method helps to elucidate effects of the aperture's boundaries on the diffraction, as demonstrated in recent studies [3, 6, 7, 8].

The diffraction being a generic wave phenomenon, interest in BDWT has risen independently in quantum mechanics. In particular, the interest in studies of sharp quantum-mechanical transients and singular wave functions, which were initiated by Moshinsky [9], was rekindled by recent progress in atom trapping [10]. In this context, it was demonstrated that wave functions with sharp transitions exhibit universal behavior [11, 12]. That is, the wave-function dynamics is predominantly determined by values of the wavefunction on its sharp boundaries. Thus, different wave functions with the same boundaries will exhibit the same dynamics. These universalities were validated in other realizations of Schrödinger-like dynamics, such as dispersion [13] and diffraction acting in the framework of the paraxial wave equation. In Refs. [14, 15], these universalities were demonstrated experimentally and were used to evaluate the diffraction of a plane wave from an aperture which can be constructed, in a "cubistic style", as a set of multiple squares.

In the present work, we develop an essential generalization of Ref. [15]. The approach is based on the possibility that, similar to the one-dimensional (1D) case, where the diffraction effect, in the near-field area, is mainly determined by the field's value at the boundaries, in the 2D case the diffracted field should be mainly determined by values of the field at the aperture's boundaries, in accordance with BDWT. In contrast to Ref. [5], the proposed formalism does not involve singularities.

The derivation consists of three stages. At the first one, we follow Ref. [13] and show that, in the case of a uniform beam, the diffracted wave is given by an expression that depends *solely* on the aperture's boundaries geometry. The second stage addresses non-uniform beams, which have uniform field values, at least, at the aperture's boundaries. To solve this problem, we follow Ref. [15] and rewrite the beam's field as a superposition of continuous and discontinuous parts. The former one experiences mild (in many cases, negligible) diffraction, while the latter part, which undergoes strong diffraction, can be evaluated exactly using the contour integral. At the third stage, we address the generic case of beams with arbitrary profiles. In this case, the previous stage's expression (with the contour integral taking into account variations of the beam's profile) is a good approximation for short distances, i.e., for the near-field area. Besides the theoretical value of this finding, the method has a potential to become an efficient tool in analyzing the beam propagation in settings which demand complex high-speed calculations, such as optical imaging of living specimens [16], beam-shape optimization [17], digital holography or computer-generated holograms [18, 19], etc.

Moreover, beyond the beam's initial boundary, i.e., in the geometrical-shadow domain, the dynamics is mainly governed by the boundaries, therefore the exact solution is solely determined by the beam's boundaries. This may lead to the design of new-edge recognition techniques in a variety of fields, such as optical eavesdropping [20] and image recognition [21].

## 2. The general derivation

The paraxial wave equation is one of the most ubiquitous approximations replacing full scalar wave equations [22], which, in turn, are valid whenever a light beam is diffracted through an aperture whose dimensions are much larger than the beam's wavelength [23]. The paraxial wave equation describes the propagation of light in homogenous and isotropic media with refractive index $n$, carrier frequency $\omega$ and wavenumber $k = \omega n / c$ ($c$ the speed of light in vacuum) as

$$\frac{\partial^2 A}{\partial x^2} + \frac{\partial^2 A}{\partial y^2} + 2ik\frac{\partial A}{\partial z} = 0 \ , \tag{1}$$

where $A(x, y, z)$ is the envelope of the electromagnetic field,

$$E(x, y, z, t) = A(x, y, z)\exp(ikz - i\omega t), \tag{2}$$

assuming that the envelope varies very slowly on the beam's wavelength scale, i.e., the amplitude obeys the paraxial approximation,

$$|\partial A / \partial z| \ll |kA| . \tag{3}$$

Accordingly, the paraxial wave equation is most commonly used to model in slowly expanding beams [24, 25]. It should be stressed that diffraction due to spatial confinement of sharp (hard-edged) boundaries does not violate the paraxial assumption [22, 15, 26, 14, 27].

The general solution to Eq. (1) is derived using the convolution (the paraxial diffraction integral) [23]

$$A(x, y; z) = \int_{-\infty}^{+\infty}\int_{-\infty}^{+\infty} A(x', y'; 0)\frac{k}{2\pi i z}\exp\left[\frac{ik}{2z}\left((x-x')^2 + (y-y')^2\right)\right]dx'dy' \tag{4}$$

where primes represent coordinates in the aperture's plane. Equation (4) with an additional phase factor, $\exp(ikz)$ is sometimes referred to as the Fresnel (paraxial) diffraction integral [15]. If the diffraction screen, with an arbitrary aperture shape, located at $z=0$, is illuminated by a field profile $a(x, y; 0)$, then the beam's profile beyond the screen can be written as (see Figure 1)

$$A(x, y; z = 0) \equiv a(x, y; z = 0)H(R(\theta) - r), \tag{5}$$

where $H$ is the Heaviside step function, and therefore $A(x,y;0)=0$ for $r=\sqrt{x^2+y^2}>R(\theta)$. Substituting (5) in (4) yields

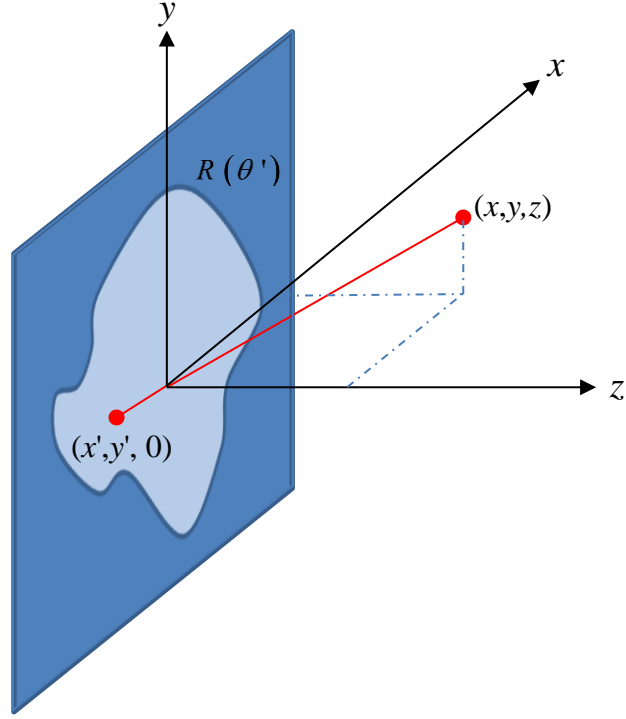

**Figure 1**: The aperture's geometry (the bright region). Coordinates in the aperture's plane are denoted by primes. The local radius of the aperture's boundary is $R(\theta')$. The transmitted field is measured at point $(x,y,z)$.

$$A(x,y;z) = \iint_{r \leq R(\theta')} a(x',y';0) \frac{k}{2\pi i z} \exp\left[\frac{ik}{2z}\left((x-x')^2 + (y-y')^2\right)\right] dx'dy' \quad (6)$$

To convert surface integral (6) into a contour integral, we define $\mathbf{F} = L(x,y;z)\hat{x} + M(x,y;z)\hat{y}$ with

$$M(x,y;z) \equiv \frac{1}{2}\sqrt{\frac{k}{2\pi i z}} a(x',y';0) \frac{1}{2} \operatorname{erfc}\left(\frac{x-x'}{\sqrt{2zik^{-1}}}\right) \exp\left(-\frac{(y-y')^2}{2zik^{-1}}\right),$$

$$L(x,y;z) \equiv -\frac{1}{2}\sqrt{\frac{k}{2\pi i z}} a(x',y';0) \frac{1}{2} \operatorname{erfc}\left(\frac{y-y'}{\sqrt{2zik^{-1}}}\right) \exp\left(-\frac{(x-x')^2}{2zik^{-1}}\right). \quad (7)$$

Then, using $\frac{d}{dz}\operatorname{erfc}(z) = -\frac{2e^{-z^2}}{\sqrt{\pi}}$, we obtain

$$\frac{\partial L}{\partial y'} = -\frac{1}{2}\sqrt{\frac{k}{2\pi i z}} \left\{ a(x',y';0) \left( \frac{k}{\sqrt{2\pi i z}} \exp\left( ik \frac{(x-x')^2 + (y-y')^2}{2z} \right) \right) + \right.$$
$$\left. \frac{\partial a(x',y';0)}{\partial y'} \frac{1}{2} \mathrm{erfc}\left( \frac{y-y'}{\sqrt{2zik^{-1}}} \right) \exp\left( -\frac{(x-x')^2}{2zik^{-1}} \right) \right\}$$

$$\frac{\partial M}{\partial x'} = \frac{1}{2}\sqrt{\frac{k}{2\pi i z}} \left\{ a(x',y';0) \left( \frac{k}{\sqrt{2\pi i z}} \exp\left( ik \frac{(x-x')^2 + (y-y')^2}{2z} \right) \right) + \right.$$
$$\left. \frac{\partial a(x',y';0)}{\partial x'} \frac{1}{2} \mathrm{erfc}\left( \frac{x-x'}{\sqrt{2zik^{-1}}} \right) \exp\left( -\frac{(y-y')^2}{2zik^{-1}} \right) \right\}.$$

(8)

Therefore, the expression for $\nabla \times \mathbf{F} = \left( \frac{\partial M}{\partial x'} - \frac{\partial L}{\partial y'} \right) \hat{z}$ amounts to the integrand of Eq.(6),

$$a(x',y';0) \frac{k}{2\pi i z} \exp\left[ \frac{ik}{2z} \left( (x-x')^2 + (y-y')^2 \right) \right], \qquad (9)$$

plus an additional term due to the derivatives of $a(x',y';0)$:

$$u(x,y;z) \equiv \frac{1}{4}\sqrt{\frac{k}{2\pi i z}} \left[ \frac{\partial a(x',y';0)}{\partial x'} \mathrm{erfc}\left( \frac{x-x'}{\sqrt{2zik^{-1}}} \right) \exp\left( -\frac{(y-y')^2}{2zik^{-1}} \right) \right.$$
$$\left. + \frac{\partial a(x',y';0)}{\partial y'} \mathrm{erfc}\left( \frac{y-y'}{\sqrt{2zik^{-1}}} \right) \exp\left( -\frac{(x-x')^2}{2zik^{-1}} \right) \right] \qquad (10)$$

Therefore, the solution is obtained as

$$A(x,y,z) = \iint_S \nabla \times \mathbf{F} \cdot d\mathbf{s}' - \iint_S u\,ds' \qquad (11)$$

Using the Stoke's theorem [5],

$$\iint_S \nabla \times \mathbf{F} \cdot d\mathbf{s}' = \oint_\Gamma \mathbf{F} \cdot d\mathbf{l}, \qquad (12)$$

where the integral surface $S$ is defined by $r' \le R(\theta')$, and contour $\Gamma$ is defined by $r' = R(\theta')$, the solution is obtained (without any approximation) as

$$A(x,y,z) = \oint_\Gamma \mathbf{F} \cdot d\mathbf{l} - \iint_S u\,ds', \qquad (13)$$

i.e., the field can be rewritten as a superposition of a contour and surface integrals, which depend on the partial derivatives of the initial profile $a(x',y';0)$. It should be stressed that a similar expression can be derived from the 2D divergence theorem [28]. Hereafter, we define

$$B(x,y;z) \equiv \oint_\Gamma \mathbf{F} \cdot d\mathbf{l} = \frac{1}{2}\sqrt{\frac{k}{2\pi i z}}$$
$$\oint_\Gamma \frac{a(x,y;0)}{2} \left\{ -\mathrm{erfc}\left( \frac{y-y'}{\sqrt{2zik^{-1}}} \right) \exp\left( ik\frac{(x-x')^2}{2z} \right) \hat{x} + \mathrm{erfc}\left( \frac{x-x'}{\sqrt{2zik^{-1}}} \right) \exp\left( ik\frac{(y-y')^2}{2z} \right) \hat{y} \right\} \cdot d\mathbf{l} \qquad (14)$$

hence $A(x,y;z) = B(x,y;z) - \iint_S u\,ds'$.

## Cylindrical coordinates

In cylindrical coordinates, one has $x = r\cos\theta, y = r\sin\theta$, $x' = R(\theta')\cos\theta', y' = R(\theta')\sin\theta'$, where the primes represent the coordinates along the aperture's boundary.

Therefore, $dx' = \left(\frac{dR(\theta')}{d\theta'}\cos\theta' - R(\theta')\sin\theta'\right)d\theta'$ and $dy' = \left(\frac{dR(\theta')}{d\theta'}\sin\theta' + R(\theta')\cos\theta'\right)d\theta'$, and then Eq.(14) can be rewritten as

$$B_a(r,\theta;z) = \frac{1}{2}\sqrt{\frac{k}{2\pi i z}} \times$$

$$\int_0^{2\pi} \frac{a(x',y';0)}{2}\left\{-\mathrm{erfc}\left(\frac{r\sin\theta - R(\theta')\sin\theta'}{\sqrt{2zik^{-1}}}\right)\exp\left(ik\frac{(r\cos\theta - R(\theta')\cos\theta')^2}{2z}\right)\left(\frac{dR(\theta')}{d\theta'}\cos\theta' - R(\theta')\sin\theta'\right)\right.$$

$$\left. +\mathrm{erfc}\left(\frac{r\cos\theta - R(\theta')\cos\theta'}{\sqrt{2zik^{-1}}}\right)\exp\left(ik\frac{(r\sin\theta - R(\theta')\sin\theta')^2}{2z}\right)\left(\frac{dR(\theta')}{d\theta'}\sin\theta' + R(\theta')\cos\theta'\right)\right\}d\theta' \quad (15)$$

In most cases, the expression given by Eq. (15) can be evaluated numerically.

## 3. The uniform field distribution

In the case where $a(x,y;0)$ amounts to constant $a_0$, the partial derivatives $\partial a/\partial x, \partial a/\partial y$ in Eq. (10) vanish and therefore, the solution is *exactly* $A(x,y,z) = B(x,y,z)$. Further, note that

$$B(x,y;z=0)/a_0 = H(R(\theta) - r). \quad (16)$$

where $H$ is again the Heaviside step function. In this case, $B(x,y;z)/a_0$ is the *exact* solution for the propagation of the initial discontinuous beam $H(R(\theta) - r; z)$. The scenario where $a(x,y;0)$ depends on $(x,y)$ will be discussed below. First, we will present some relevant examples of uniform beams.

**Example 1. The analytical solution for the edge**

Consider the case of an edge infinite in the y-direction,

$$A(x,y;0) = a_0 H(-x). \quad (17)$$

In this case,

$$A(x,y;z) = B(x,y,z) = \frac{a_0}{4}\sqrt{\frac{k}{2\pi i z}}\mathrm{erfc}\left(\frac{x}{\sqrt{2zik^{-1}}}\right)\int_{-\infty}^{\infty}\exp\left(ik\frac{(y-y')^2}{2z}\right)dy'$$

$$= \frac{a_0}{2}\mathrm{erfc}\left(\frac{x}{\sqrt{2zik^{-1}}}\right). \quad (18)$$

This solution is clearly consistent with the 1D case (see Ref. [9]).

**Example 2. The analytical solution for a rectangular aperture**

Considering the aperture with a rectangular shape, i.e.,

$$A(x,y;0) = a_0 \left[ H(x+L) - H(x-L) \right]\left[ H(y+L) - H(y-L) \right], \quad (19)$$

One obtains

$$A(x,y;z) = B(x,y;z) = \frac{a_0}{4}\sqrt{\frac{k}{2\pi i z}} \times$$

$$\oint_\Gamma \left\{ \text{erfc}\left(\frac{x-x'}{\sqrt{2zik^{-1}}}\right) \exp\left(ik\frac{(y-y')^2}{2z}\right)\hat{y} - \text{erfc}\left(\frac{y-y'}{\sqrt{2zik^{-1}}}\right) \exp\left(ik\frac{(x-x')^2}{2z}\right)\hat{x} \right\} \cdot d\mathbf{l} = \quad (20)$$

$$a_0 \left\{ \frac{1}{2}\left[ \text{erfc}\left(\frac{x+L}{\sqrt{2zik^{-1}}}\right) - \text{erfc}\left(\frac{x-L}{\sqrt{2zik^{-1}}}\right) \right] \frac{1}{2}\left[ \text{erfc}\left(\frac{y+L}{\sqrt{2zik^{-1}}}\right) - \text{erfc}\left(\frac{y-L}{\sqrt{2zik^{-1}}}\right) \right] \right\}.$$

Therefore, the result is a product of two smooth rectangular functions in accordance with Ref. [15]. This example illustrates the fact that Eq. (15) is valid even when boundaries of the aperture have zero transition width (in Eq. (36) of Ref. [15], the boundaries were smooth with a *nonzero* transition width).

**Example 3. The numerical solution for a circular aperture**

Figure 2 shows a comparison between results produced by the contour integration, i.e. Eq.(15), and the 2D surface integration, i.e., Eq.(4), for a circular aperture at $z = 70\text{mm}$. As can be seen from the lower right panel of the figure (which represents the amplitude's cross-section along the *x*-axis), the analytical solution, i.e., Eq.(15) (the solid red curve) is in perfect agreement with the numerical solution (the dashed black curve).

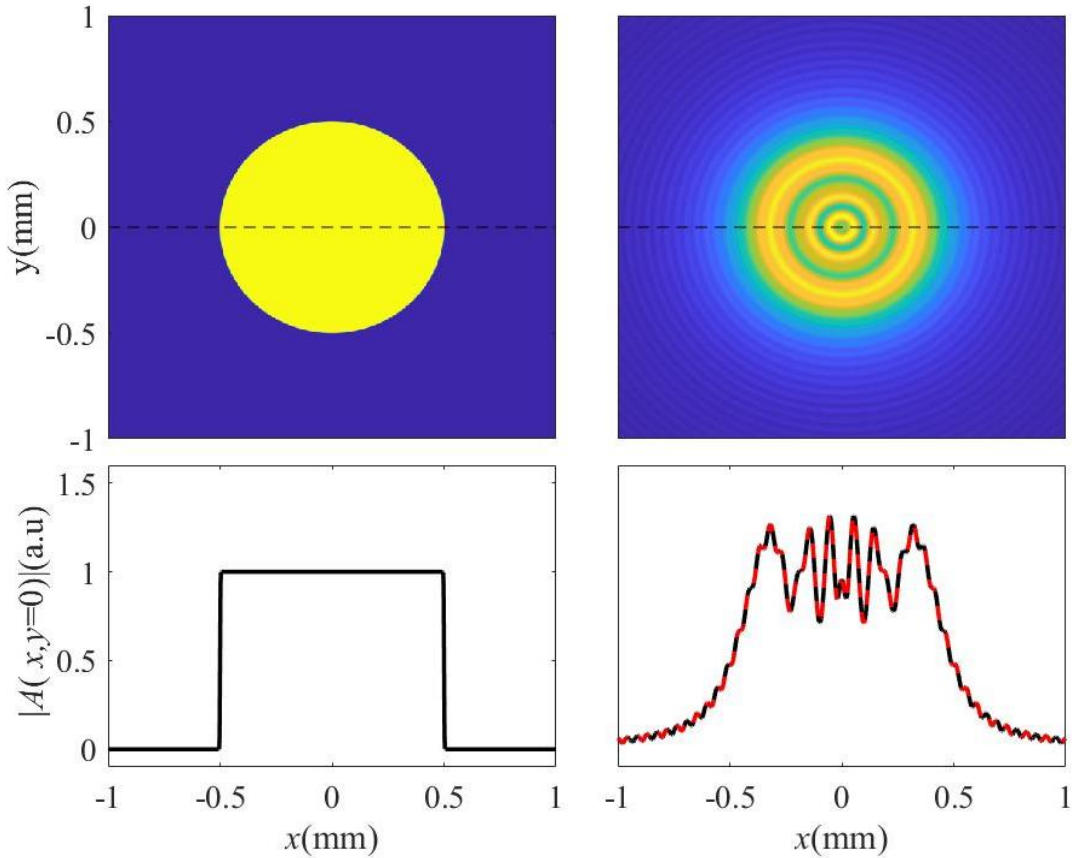

**Figure 2** – Diffraction of a uniform beam's profile via a circular aperture - comparison between Eq. (15) and the surface integration produced by Eq.(4). The upper-left panel presents the initial beam's amplitude profile, while the lower-left panel shows its cross-section along the *x*-axis (the cross-section is marked by the dashed black curve in the upper left-panel). The upper-right panel is the final numerically exact amplitude at $z = 70\text{mm}$, while the lower-right panel shows its cross-section along the *x*-axis. The solid black curve represents the exact numerical calculation produced by Eq.(4), and the dashed red curve corresponds to Eq. (15). In this case, the beam's carrier wavelength is 628 nm, and the aperture's radius is 0.5 mm.

**Example 4. The numerical solution for a complex aperture**

Figure 3 displays the comparison between the results of the contour integration [Eq. (15)] and the 2D surface integration [Eq. (4)], which was applied to the aperture presented by

$$R(\theta') = R_0 \left(1 - 0.5\cos(4\theta')\right) \qquad (21)$$

with $R_0 = 0.5\text{mm}$. As can be seen in the figure, the agreement between the two methods is perfect. The numerical exact solution (the dashed black line) and the analytical expression, given by Eq. (15) for the aperture's shape (21), (the solid red line) are practically identical. These results support our above-mentioned finding, which states that Eq. (15) is the exact solution in the case of a uniform field distribution, regardless of the aperture's boundaries shape.

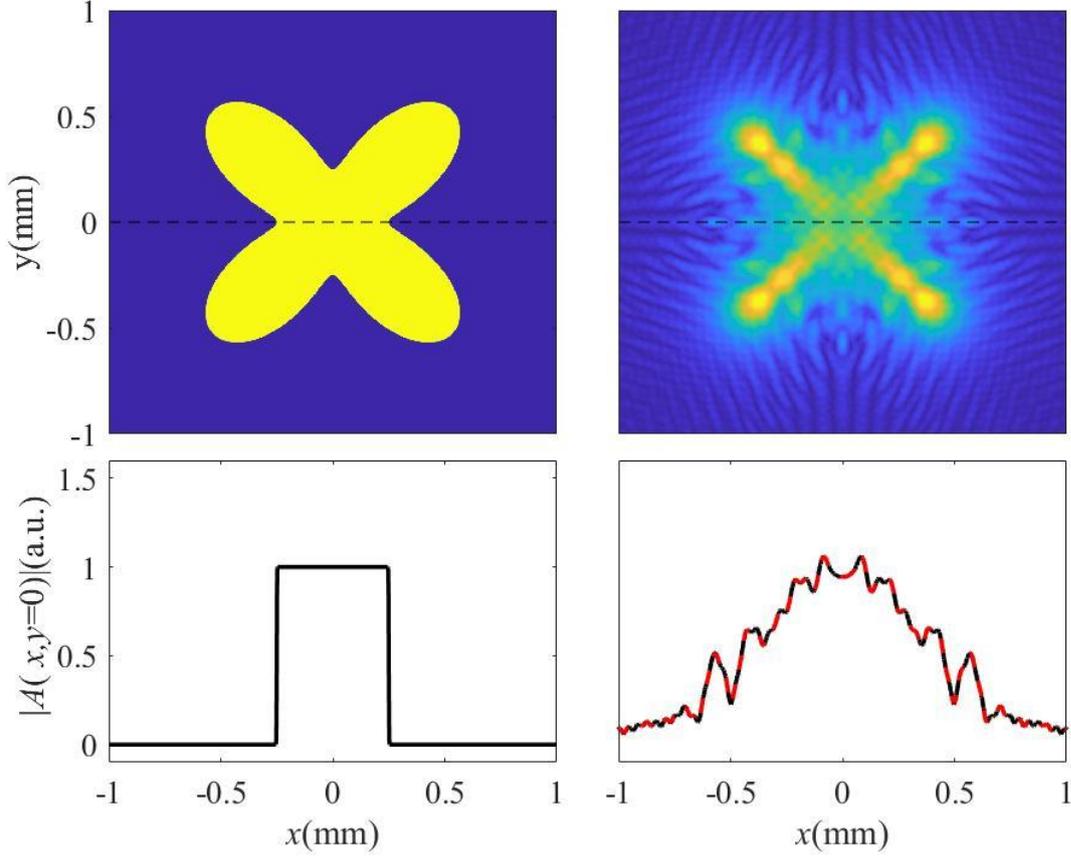

**Figure 3** – The same as in Fig. 2 (i.e., the same arrangement of panels, and the field is uniform in the aperture), but for the aperture defined as per Eq. (21). As in Fig. 2, the dashed red and solid black curves, in the lower-right panel, represent solutions given by Eqs. (15) and (4) at $(x, 0, z = 70\text{mm})$, respectively (i.e., along the cross-section marked by the dashed black curve in the upper right panel).

## 4. The general field distribution with a uniform value over the aperture's boundary

In cases where the initial beam's field $a(x, y; 0)$ varies in space *but remains constant over the contour*, one has the situation with $A(x, y; z) \neq B(x, y; z)$, and even $A(x, y; 0) \neq B(x, y; 0)$.

However, the difference $A(x, y; 0) - B(x, y; 0)$ vanishes on the aperture's boundary (see, for example, the lower-left panel of Figure 4). Therefore, we can follow the lines of Ref. [21] to rewrite the initial beam's profile as

$$A(x, y; 0) = \left[A(x, y; 0) - B(x, y; 0)\right] + B(x, y; 0) \,. \qquad (22)$$

The term in the square brackets is continuous; hence, it undergoes much milder evolution than the singular part $B(x,y;0)$. As $B(x,y;0)$ is known *exactly*, according to Eq.(14), the propagation in the near-field area takes the form of (for $z=0$)

$$A(x,y;z) \approx \left[A(x,y;0) - B(x,y;0)\right] + B(x,y;z) . \qquad (23)$$

It should be emphasized that by no means $A(x,y;0) - B(x,y;0)$ is necessarily smaller than $B(x,y;z)$. In fact, it can be even larger. However, the effect of diffraction on the former part is negligible, on short distances, in comparison to its effect on the latter one. In Figure 4, a comparison between Eq. (23) and the exact numerical calculation is presented for the same aperture's boundary as defined by Eq. (17), but with the amplitude distribution

$$a(\theta,r;0) = 1 + \exp\left(-(r/R(\theta))^2\right)/2 . \qquad (24)$$

In the lower-left panel of Fig.4 the solid black curve represents initial values of the amplitude in the cross-section, i.e., $A(x,0;0)$. The dashed-red curve represents initial expression $B(x,0;0)$, and the dotted blue curve represents difference $A(x,0;0) - B(x,0;0)$, whose diffraction effects are inconspicuous. The figure demonstrates excellent agreement between the two results, given by Eqs. (15) and (4) (see the lower right panel), despite the beam's non-uniformity. Although the results are not identical, especially in the vicinity of the boundaries, the difference is insignificant.

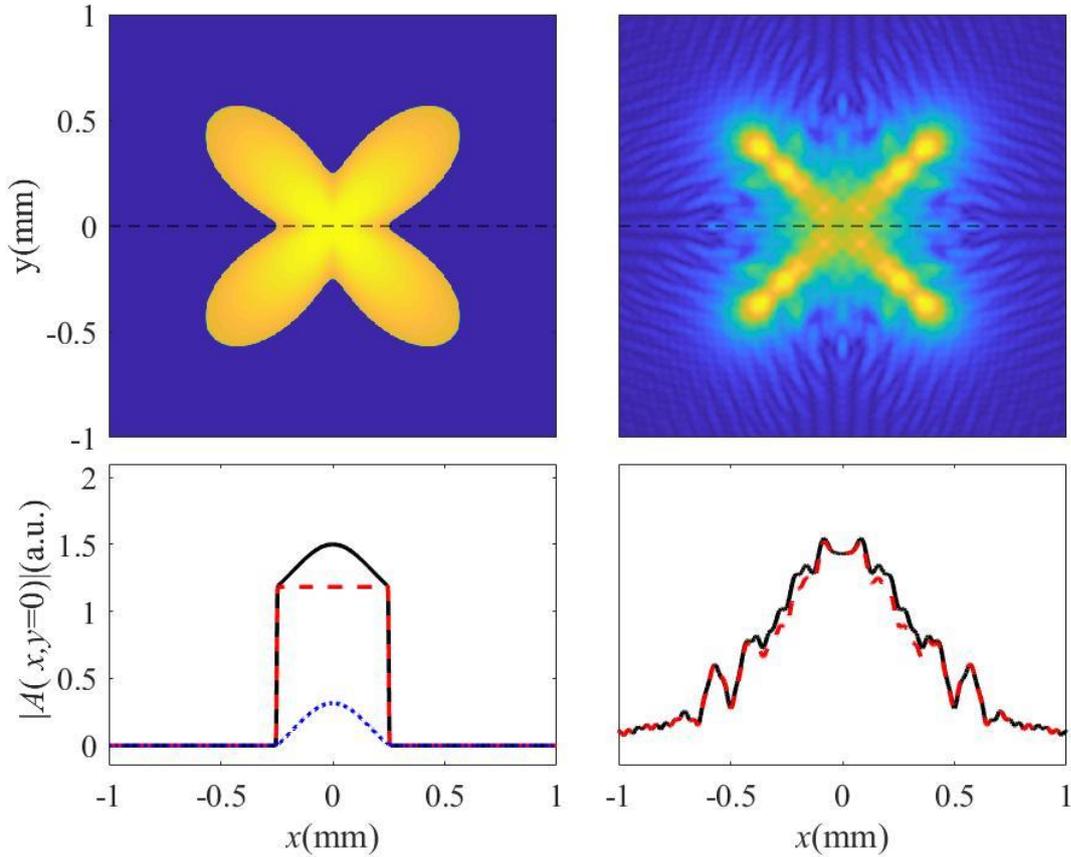

**Figure 4** – The same aperture's shape and panel arrangement as in Figure 3, but for a beam with the profile defined as per Eq.(24). The black solid curve in the lower-left panel shows the cross section of the initial profile along the *x*-axis, while the dashed red and dotted blue curves represent the profiles' cross sections $B(x, y=0; z=0)$ and $A(x, y=0; z=0) - B(x, y=0; z=0)$, respectively. The upper-right panel displays the beam at $z=70\text{mm}$. In the lower-right panel, the dashed red and solid black curves represent Eqs.(23) and (4), at $(x,0;z=70\text{mm})$ respectively.

# 5. Arbitrary field distribution within an arbitrary aperture shape

In the case when the beam's field varies over the aperture's boundaries, the term $A(x,y;0)-B(x,y;0)$ *does not* vanish on the aperture's boundaries. Nevertheless, this term is *continuous* everywhere (including the boundaries – see, for example, the lower-left panel of Figure 6), hence we can follow the lines of the previous section, rewriting the initial field like in Eq.(22). Hence, due to the continuity of $A(x,y;0)-B(x,y;0)$, it is less affected by the diffraction, again yielding Eq.(23) in the near-field area, i.e., for $z>0$,

$$A(x,y;z) \approx A(x,y;0) + \left[ B(x,y;z) - B(x,y;0) \right] \qquad (25)$$

Figure 5 and Figure 6 display the comparison between Eq. (23) and the 2D paraxial diffraction integral, for the case when the initial field distribution is

$$a(x,y;0) = 1 + x/2, \qquad (26)$$

within a circular aperture, for $z=70\,\text{mm}$ and $z=400\,\text{mm}$ respectively. Figure 7 and Figure 8 address the same field distribution, for $z=70\,\text{mm}$ and $z=10\,\text{mm}$, respectively, but for the aperture presented by Eq.(21). These examples show that Eq.(23) agrees well with the exact solution.

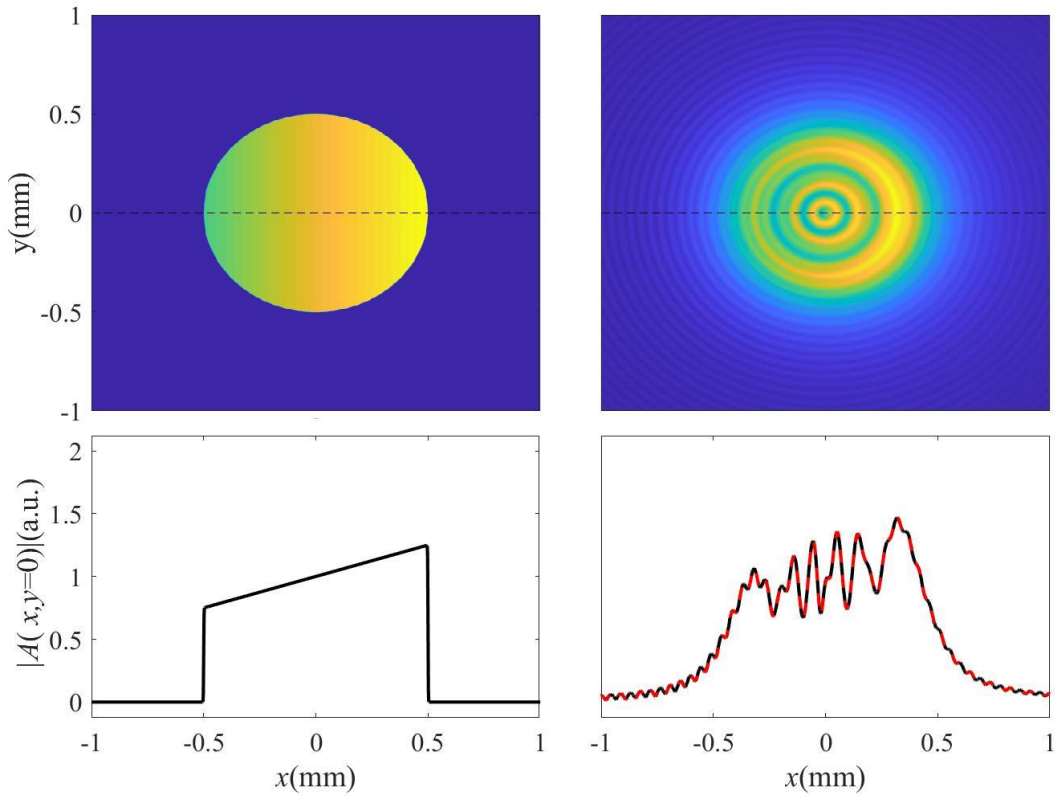

**Figure 5.** The same as in Fig.2, i.e., for the circular aperture, but with the non-uniform amplitude profile determined by Eq. (26). The left and right columns stand for $z=0$ and $z=70\,\text{mm}$, respectively. The lower panel represents cross sections $(x, y=0)$ of the upper ones. In the lower-right panel, the solid black curve represents the exact numerical calculation produced by Eq.(4), and the dashed red curve represents the result given by Eq. (25).

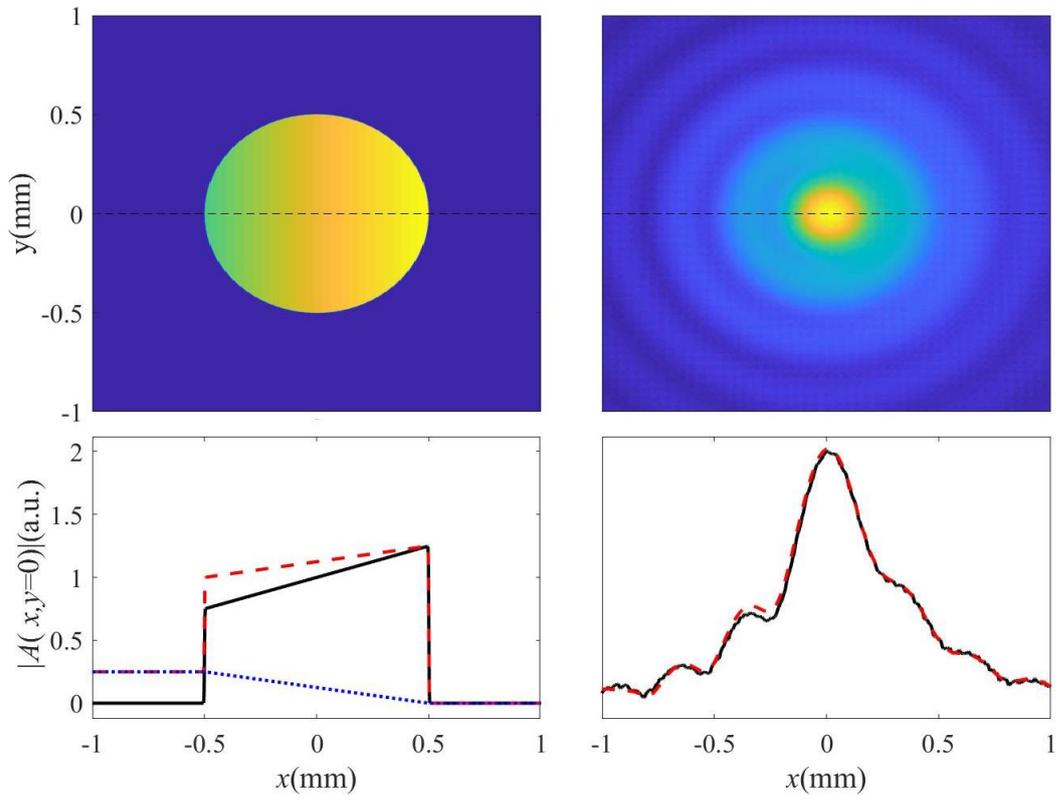

**Figure 6.** The same aperture and beam's profile as in Figure 5, but for $z = 400\,\text{mm}$ in the right panels. In the lower-left panel, the dashed red and dotted blue curves represent the profile cross-sections of $B(x,0;0)$ and $A(x,0;0) - B(x,0;0)$, respectively. The solid black curve shows the cross-section of the initial profile along the $x$-axis. In the lower-right panel, the solid black curve represents the exact numerical calculation produced by Eq. (4), and the dashed red curve represents the result given by Eq. (25) (both in the $x$-axis cross-section).

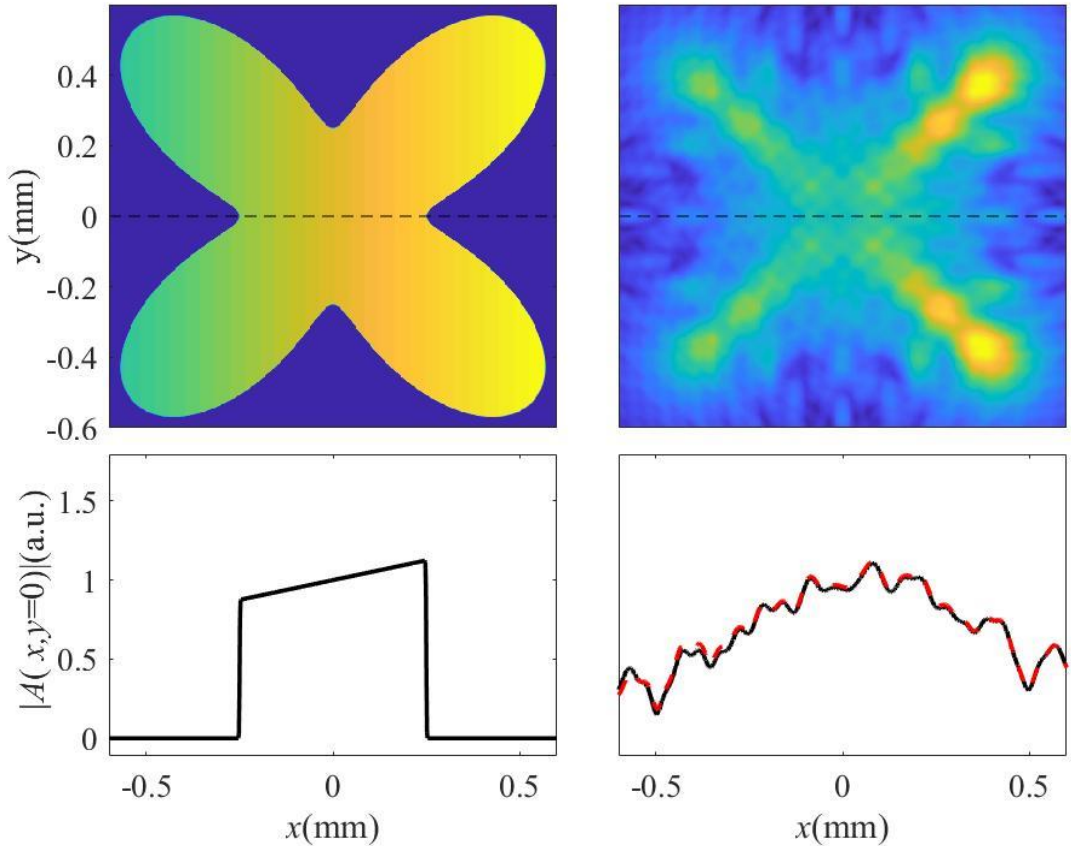

**Figure 7** – Comparison between the result given by Eq. (25) and the exact numerical solution [Eq.(4)] for an aperture, whose boundaries are determined by Eq. (21) (as in Fig.3), while the amplitude of the beam's profile is determined by Eq. (26). The left and right columns pertain to $z = 0$ and $z = 70\,\text{mm}$, respectively. In the lower-right panel, the solid black curve represents the exact numerical calculation produced by Eq. (4), and the dashed red curve represents the result given by Eq. (25) (both in the $x$-axis cross-section).

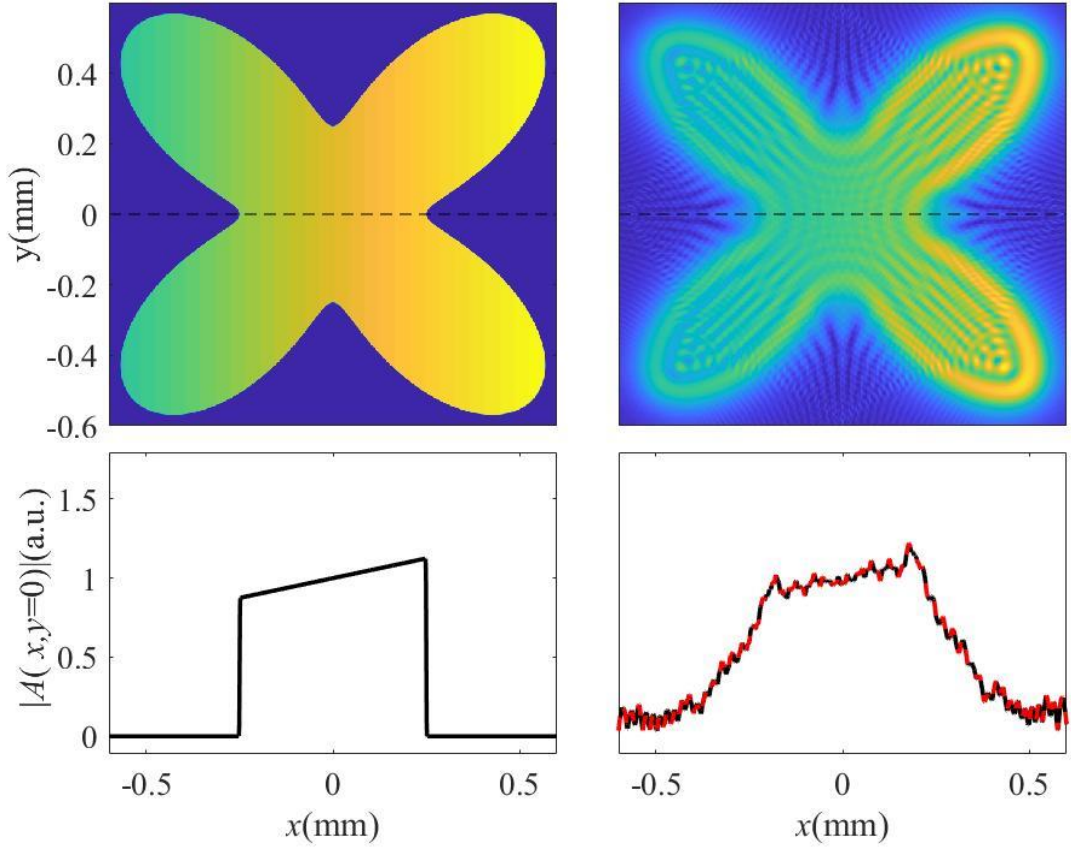

**Figure 8** – same as in Figure 7, but for propagation distance $z = 10\,\text{mm}$.

The accuracy range of the approximation depends on the smoothness of $A(x,y;0) - B(x,y;0)$. The smoother is $A(x,y;0) - B(x,y;0)$, the wider the accuracy range will be. For example, if $a(x,y;0)$ in Eq. (23) is constant, it follows that $A(x,y;0) - B(x,y;0) = 0$, hence, $A(x,y;z) = B(x,y;z)$ in accordance with Eqs. (13) and (14).

These findings emphasize the fact that the change in the near field, i.e., $B(x,y;z) - B(x,y;0)$, is mostly governed by the boundaries. Moreover, as can be seen from Figure 4, Eq. (23) provides a better approximation at the beam's periphery.

These results attest to the universality of the method presented in Refs. [15, 14, 13], but now in the 2D setting. Thus, there always exists a regime in which the diffraction effect depends, primarily, on the shape of the aperture's boundaries and values of the incident field at the boundaries.

Thus far, we addressed the setting with the single aperture. When there are multiple apertures in the diffraction screen, such as in *photon sieves*, which have recently drawn much interest, in particular, as tools for manipulations of the X-ray radiation (see, for example, Refs. [29, 30]), the above method can easily be generalized. That is, if $A_n(x,y;0)$ is the initial beam's profile of the $n$-th aperture, and $B_n(x,y;0)$ is the corresponding contour integral, then Eq. (25) can be generalized as

$$A_n(x,y;z) \cong \sum_n \left\{ A_n(x,y;0) + \left[ B_n(x,y;z) - B_n(x,y;0) \right] \right\} \qquad (27)$$

where summation is taken over all apertures in the diffraction screen.

# 6. Comparison between the new method and BDWT

In this section, we compare the new method, based on Eq. (25), to results produced by BDWT [5] (recall this acronym stands for the boundary diffraction wave theory).

## The uniform distribution

Figure 9 displays the comparison between the two methods for the uniform beam, which is diffracted by the circular aperture. In the case of a uniform distribution, the BDWT is a numerically exact form of the Kirchhoff integral, while Eq. (15) is a corollary of the paraxial wave equation. Therefore, the BDWT is expected to be more accurate at wavelength-scale distances. However, since the paraxial approximation is accurate in the vicinity of the boundaries, and since, due to the singular nature of the boundaries, most of the distortions occur there, the method developed above, which is based on the paraxial approximation, reveals high accuracy in comparison to the BDWT.

Nevertheless, there are two noteworthy differences. The main difference between the two methods appears at the center of the aperture (the farthest point from the boundaries – see the left-column plots in Fig. 9), due to constructive interferences of the radiation emitted from the aperture's boundaries. Furthermore, the BDWT integration includes a singularity point, which can be seen in the lower-right panel of Fig. 9. However, besides these two differences (actually, minor ones), the two methods (BDWT and Eq. (15)) demonstrate excellent agreement with the numerical solution produced by Eq. (4).

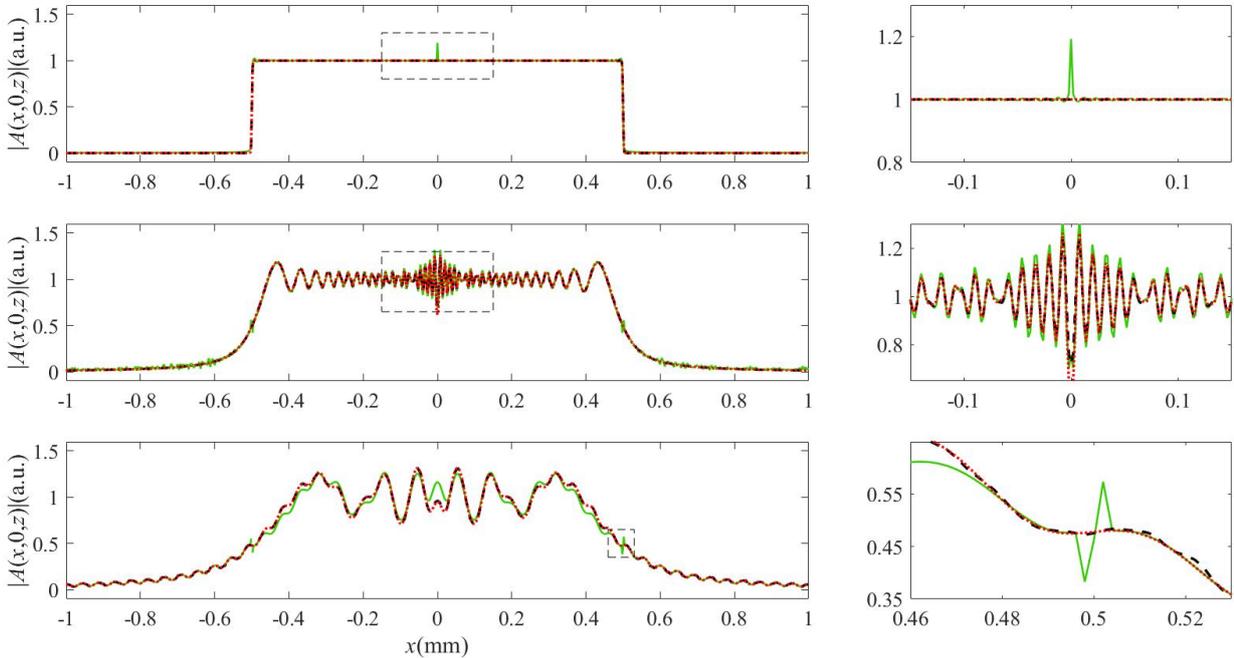

**Figure 9** – Comparison between the BDWT and the method given by Eq. (15) for the diffraction of the uniform beam's profile through the circular aperture (the same aperture's shape and initial field distribution as in Figure 2). The plots represent numerically found value of the electric field at $(x, y = 0; z)$. Results produced by Eq. (15) are marked by red dots; those generated by the surface integration in Eq.(4) are marked by the dashed black curve; and results of the BDWT method of Ref. [5] are denoted by the green solid curve. The upper, center, and lower panels represent results for $z = 2\lambda \approx 1.2\,\mu\text{m}, 10\,\text{mm}, \text{and } 70\,\text{mm}$, respectively. Panels in the right column zooms small dashed boxes from the left column.

## The non-uniform distribution

To illustrate the scenario initiated by non-uniform inputs, we follow Ref. [31] and apply the paraxial approximation to a Gaussian beam [6]. Figure 10 presents comparison between Eq. (23), the BDWT, and the numerical solution based on Eq.(4). Since all three methods are based on the paraxial approximation, the disagreement at the center of the aperture is negligible, especially for short distances ($z \cong \lambda$). However, the advantage of the BDWT method becomes apparent for longer distances, since the new method is designed for the short-distance (near field) range. Nevertheless, it is seen in Fig. 10 that the accuracy is excellent even for much longer distances. It should be noted that even in the paraxial approximation the singularities of the BDWT method have to be attended (see plots in the right column of Fig.10).

To conclude this section, the new method elaborated above presents, in spite of its simplicity, very accurate agreement with the results produced by the BDWT method.

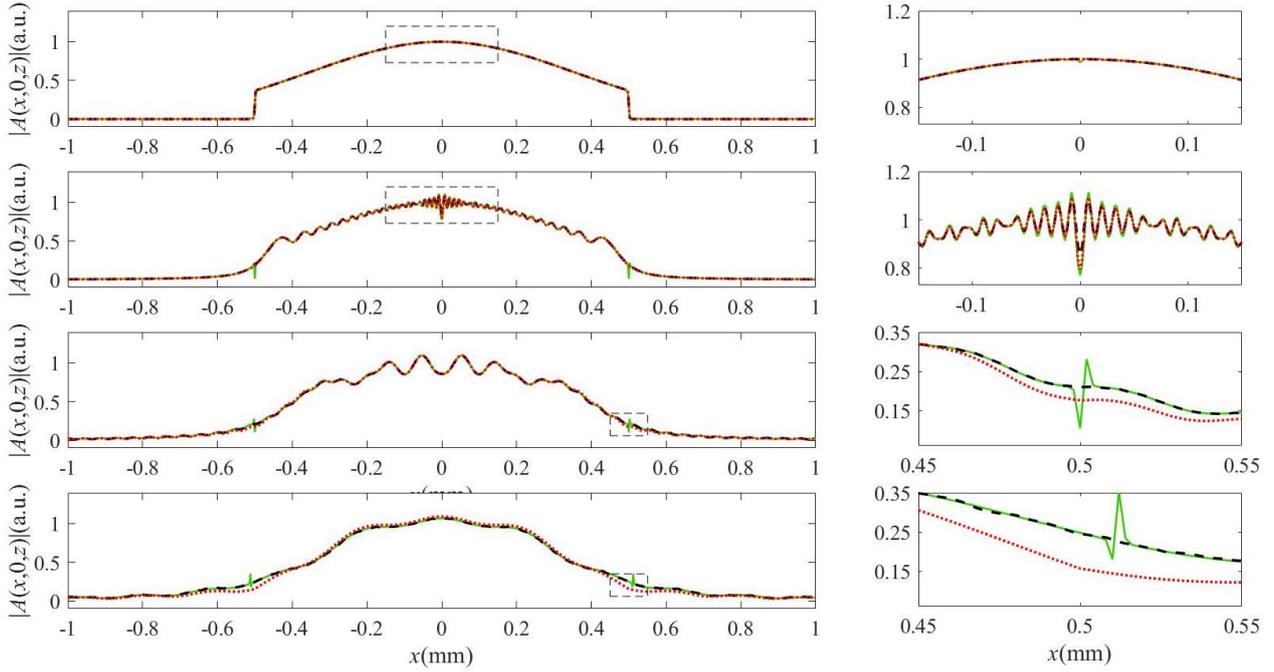

**Figure 10** –Comparison between results produced by the BDWT method (as given in Refs. [6] and [31]) and by Eq. (23) for the diffraction of the beam with the Gaussian profile through the circular aperture (the same aperture as in Figure 2). The plots display numerical values of the electric field at $(x, y = 0; z)$, where $z = 2\lambda \sim 1.2\,\mu\text{m}, 10\,\text{mm}, 70\,\text{mm}, \text{and } 270\,\text{mm}$, from the upper panel to the lower one, respectively. Results produced by Eq. (23) are marked by red dots; those obtained from the surface integration in Eq.(4) are marked by the dashed black curve; and the results generated by the BDWT method are denoted by the green solid curve. Panels in the right column zoom small dashed boxes in the left column. The results pertain to the following parameters: the waist of the smallest spot size is $w_0 = 0.5\,\text{mm}$ (the aperture's radius), with $w_0$ taken in the aperture plain $(z = 0)$, and the wavenumber is $k = 10^7\,\text{m}^{-1}$.

## Summary and Conclusion

This paper continues the line of research elaborated by Miyamoto and Wolf [5]. Accordingly, the propagation of the diffracted beam is separated into two components: the first one is determined by aperture's boundaries, while the second field component is independent of them. The difference between the present analysis from that developed in Ref. [5], which dissects space into different domains, is that the present method provides a solution that is well defined in the entire space. The corresponding expression is simpler than in Ref. [5], and in some cases, it can be obtained in the analytical form. Moreover, unlike the method given in Ref. [5], the presented one does not include singularities at the boundary of the geometrical shadow.

Similar to the 1D solution elaborated in Ref. [15], our method emphasizes the universality of the diffraction effects. In particular, a manifestation of the universality is that, in the near field, the effects are chiefly determined by values of the field at boundaries of the aperture.

It is shown that, in the case of an incident plane wave, the method yields exact solutions (within the paraxial approximation), regardless of the aperture's shape. Moreover, even when the beam's power is distributed non-uniformly across the aperture, the method yields accurate agreement with the numerically exact solution in the near-field area.

E.G. would like to thank Avi Marchewka for interesting discussions.

E.L. would like to dedicate this article to the memory of Zohar Ofri.
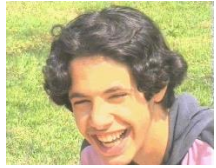

Zohar Ofri

## References

[1]  M. Born and E. Wolf, Principles of Optics, 1980 Oxford: Pergamon, Section 8.9.

[2]  E. Hecht, Optics, 2002 Addison Wesley.

[3]  R. Kumer, S. K. Kaura, D. P. Chhachhia and A. K. Aggarwal, Direct visualization of Young's Boundary Diffraction Wave, 2007 Optics Communications, 276, 54–57.

[4]  A. Rubinowicz, Thomas Young and the theory of diffraction, 1957 Nature, 180, 160.

[5]  K. Miyamoto and E. Wolf, Generalization of the Maggi-Rubinowicz Theory of the Boundary Diffraction Wave — Part I and II, 1962 J. Opt. Soc. Am., 52 ,615-637.

[6]  P. Piksarv, P. Bowlan, M. Lohmus, H. Valtna-Lukner, R. Trebino and P. Saari, Diffraction of Ultrashort Gaussian Pulses within the Framework of Boundary Diffraction Wave Theory, 2012 J. Opt., 14 015701.

[7]  E. Cady, Boundary diffraction wave integrals for diffraction modeling of external occulters, 2012 Opt. Express, 20-14, 15196–15208.

[8]  R. Kumar, Structure of boundary diffraction wave revisited, 2008 Appl. Phys. B.

[9]  M. Moshinsky, Diffraction in time, 1952 Phys. Rev, 88, 625-631.

[10] A. del Campo, G. Garcia-Calderon and J. G. Muga, Quantum transients, 2009 Phys. Rep, 476, 1-50.

[11] E. Granot and A. Marchewka, Generic short-time propagation of sharp-boudaries wave packets, 2005 Europhys. Lett., 72, 341-347.

[12] E. Granot and A. Marchewka, Emergence of currents as transient quantum effect in nonequilibrium systems, 2011 Phys. Rev. A, 84, 032110.

[13] E. Granot, E. Luz and A. Marchewka, Generic pattern formation of sharp-boundaries pulses propagation in dispersive media, 2012 J. Opt. Soc. Am. B, 29, 763-768.

[14] E. Luz, T. Ben Yaakov, S. Leiman, S. Sternklar and E. Granot, Generic propagation of beams with sharp spatial boundaries, 2015 J. Opt. Soc. Am. A, 32, 678-684.

[15] E. Luz, E. Granot and B. A. Malomed, Analytical solutions for beams passing apertures with sharp boundaries, 2016 J. Opt., 18.

[16] L. Gao, L. Shao, C. D. Higgins, J. S. Poulton, M. Peifer, M. W. Davidson, X. Wu, B. Goldstein and E. Betzig, Noninvasive Imaging beyond the Diffraction Limit of 3D Dynamics in Thickly Fluorescent Specimens, 2012 Cell, 151-6, 1370–1385.